\documentclass[fleqn,usenatbib]{mnras}

\usepackage{newtxtext,newtxmath}

\usepackage[T1]{fontenc}
\usepackage{ae,aecompl}

\usepackage{graphicx}	
\usepackage{amsmath}	
\usepackage{amssymb}	
\usepackage[flushleft]{threeparttable}
\usepackage[normalem]{ulem}

\newcommand{\hmsun}{h^{-1}{\rm M}_\odot}
\newcommand{\hmpc}{h^{-1}{\rm Mpc}}

\title[Non-fiducial cosmological test with voids]{
Non-fiducial cosmological test from geometrical and dynamical distortions around voids
}

\author[C. M. Correa et al.]{
Carlos M. Correa,$^{1,2}$\thanks{E-mail: cmcorrea@oac.unc.edu.ar (CMC)}
Dante J. Paz,$^{1,2}$
Nelson D. Padilla,$^{3,4}$
Andr\'es N. Ruiz,$^{1,2}$
\newauthor Ra\'ul E. Angulo,$^{5,6}$
and Ariel G. S\'anchez$^{7}$
\\
$^{1}$Instituto de Astronom\'ia Te\'orica y Experimental, UNC-CONICET, Laprida 854, X5000BGR-C\'ordoba, Argentina\\
$^{2}$Observatorio Astron\'omico de C\'ordoba, Universidad Nacional de C\'ordoba, Laprida 854, X5000BGR-C\'ordoba, Argentina\\
$^{3}$Instituto de Astrof\'isica, Pontificia Universidad Cat\'olica de Chile, Av. Vicu\~na Mackenna 4860, Santiago, Chile\\
$^{4}$Centro de Astro-Ingenier\'ia, Pontificia Universidad Cat\'olica de Chile, Av. Vicu\~na Mackenna 4860, Santiago, Chile\\
$^{5}$Donostia International Physics Centre (DIPC), Paseo Manuel de Lardizabal 4, 20018 Donostia-San Sebastian, Spain\\
$^{6}$IKERBASQUE, Basque Foundation for Science, E-48013, Bilbao, Spain\\
$^{7}$Max-Planck-Institut f\"ur extraterrestrische Physik, Postfach 1312, Giessenbachstr., 85741 Garching, Germany
}

\date{Accepted XXX. Received YYY; in original form ZZZ}

\pubyear{2018}

\begin{document}
\label{firstpage}
\pagerange{\pageref{firstpage}--\pageref{lastpage}}
\maketitle


\begin{abstract}
We present a new cosmological test using the distribution of galaxies around cosmic voids without assuming a fiducial cosmology.
The test is based on a physical model for the void-galaxy cross-correlation function projected along and perpendicular to the line of sight.
We treat correlations in terms of void-centric angular distances and redshift differences between void-galaxy pairs, hence it is not necessary to assume a fiducial cosmology.
This model reproduces the coupled dynamical (Kaiser effect, RSD) and geometrical (Alcock-Paczynski effect, GD) distortions that affect the correlation measurements.
It also takes into account the scale mixing due to the projection ranges in both directions.
The model is general, so it can be applied to an arbitrary cylindrical binning scheme, not only in the case of the projected correlations.
It primarily depends on two cosmological parameters: $\Omega_m$, the matter fraction of the Universe today (sensitive to GD), and $\beta$, the ratio between the growth rate factor of density perturbations and the tracer bias (sensitive to RSD).
In the context of the new generation of galaxy spectroscopic surveys, we calibrated the test using the Millennium XXL simulation for different redshifts.
The method successfully recovers the cosmological parameters.
We studied the effect of measuring with different projection ranges, finding robust results up to wide ranges.
The resulting data covariance matrices are relatively small, which reduces the noise in the Gaussian likelihood analysis and will allow the usage of a smaller number of mock catalogues.
The performance evaluated in this work indicates that the developed method is a promising test to be applied on real data.
\end{abstract}



\section{Introduction}
\label{sec:intro}

One of the major challenges of modern cosmology is to understand the nature of dark energy, which drives cosmic acceleration.
There is a wide variety of dark energy models, hence, in order to constrain them, it is important to apply several complementary statistical methods to the available simulated and observational data.

Cosmic voids are the subdense regions of the Universe.
Since their discovery \citep{voids_gregory_thompson,voids_kirshner,voids_lapparent}, voids have been recognized as powerful cosmological laboratories, specially with the advent of modern galaxy redshift surveys.
As they take up most of the volume of the space, they constitute a fundamental component of the cosmic web, holding valuable clues about the geometry and expansion history of the Universe.
The study of voids offers two distinct advantages over the high density regime: i) void dynamics is less nonlinear since it is mainly composed of single streaming flows, hence, it is easier to model systematics such us redshift-space distortions effects \citep{voids_sheth,voids_padilla,voids_ceccarelli,clues1,clues2,voids_hamaus,ap1_hamaus,clues3}; and ii) theories of modified gravity predict deviations from General Relativity (GR) to be most pronounced in unscreened low-density environments, making voids a powerful tool for detecting them \citep{modgr_clifton,modgr_li,modgr_clampitt,modgr_cai,modgr_barreira,modgr_lam,modgr_joyce,modgr_koyama,modgr_cautun,modgr_paillas}.

Voids can be used to infer cosmological information in different ways.
For instance, a shape analysis of stacked voids allows to perform an \citet[AP]{ap} test.
The AP test is a purely geometric method that examines the ratio of the observed angular to radial sizes of objects that are known to be intrinsically isotropic.
Many studies on this topic can be found in the literature: \citet{apvoids_ryden,apvoids_lavaux,apvoids_sutter,apvoids_mao}.
Also, cosmology can be constrained by the analysis of the abundance of voids: \citet{voids_sheth,abundance_furlanetto,abundance_jennings,abundance_chan,abundance_achitouv,abundance_pisani}.

This work is focused on another method: the void-galaxy cross-correlation function on redshift-space \citep{clues2,ap3_hamaus,ap1_hamaus,ap2_hamaus,rsd_hamaus,rsd_cai,rsd_achitouv1,rsd_achitouv2,rsd_chuang,rsd_hawken,rsd_nadathur,voidid_nadathur}.
This statistical function is a powerful tool that describes the void environment and dynamics.
Specifically, it quantifies the probability excess of having a galaxy around a void.
According to the cosmological principle, this function must be isotropic.
However, anisotropies arise due to the presence of dynamical and geometrical distortions.
Dynamical distortions, or redshift-space distortions, arise from the contribution of the line-of-sight component of the peculiar velocities of the galaxies surrounding voids.
Geometrical distortions, on the other hand, arise when a wrong cosmology is used to assign a distance scale to measure correlations.
This is a manifestation of the AP effect and can be used to extract cosmological information.
Both types of distortions are coupled, hence, any cosmological analysis must take both into account.  

In this work, in addition, we consider a third type of systematics that affects the cosmological inference when modelling the correlation function.
Models evaluate the correlation function on a given point of the space.
However, when measuring it, a binning scheme is used, and hence, several scales are mixed in the observation.
This correction is a key aspect for the new method presented here: we measure and model two fully projected void-galaxy correlation functions, along and perpendicular to the line of sight.
The data covariance matrices associated to this method are smaller, and the noise in the Gaussian likelihood analysis is reduced, which will allow the usage of a smaller number of mock catalogues \citep{Taylor13,Dodelson13}.
The other important aspect of our analysis is the treatment of correlations directly in terms of void-centric angular distances and redshift differences between void-galaxy pairs, so that it is not necessary to assume a fiducial cosmology.
The physical model that we developed reproduces all these type of distortions and depends primarily on two cosmological parameters: the matter fraction of the Universe today, $\Omega_m$, and the ratio between the growth rate of density perturbations and the linear tracer bias, $\beta$.

This paper is organised as follows.
In Section~\ref{sec:fundaments}, we explain the fundaments of the cosmological test, namely, the origin of geometrical and dynamical distortions, and the role of the projected void-galaxy cross-correlation functions as cosmological tools.
In Section~\ref{sec:data}, we provide the data sets.
We describe the numerical N-body simulation and the void catalogues.
In Section~\ref{sec:measurements}, we present the projected correlation functions from data.
In Section~\ref{sec:model}, we present a physical model for them.
In Section~\ref{sec:mcmc_fit}, we perform a likelihood analysis to constrain the cosmological parameters from the model.
Finally, we summarize and discuss our results in Section~\ref{sec:conclusions}.


\section{Fundaments of the test}
\label{sec:fundaments}

\subsection{Geometrical distortions}
\label{subsec:gd}

Our method relies on measuring the cross-correlation function between void-galaxy pairs on a spectroscopic survey.
The observables we have are $(\theta, z', z)$, where $\theta$ denotes the angular distance subtended by a void centre and a galaxy on the plane of the sky (hereafter POS), $z'$ the redshift of the void centre provided by the void finder (see Section~\ref{subsec:voids}), and $z$ the redshift of the galaxy.
Their respective POS and line-of-sight (hereafter LOS) comoving separations, $(\sigma, \pi)$, are given by the following equations: 
\begin{equation}
    \sigma = d_\mathrm{com}(z') ~ \theta \\
    \pi    = \left|d_\mathrm{com}(z)-d_\mathrm{com}(z')\right|,
	\label{eq:distance}
\end{equation}
where $d_{\rm com}$ is the comoving distance from the observer.
Analytical expressions for the comoving distance in the general non-flat case are given in terms of elliptic functions by \citet{cosmo_rollin}.
Throughout this work, we adopted a standard flat $\Lambda$CDM cosmology, but the method can be generalised to incorporate other models.
In this frame,
\begin{equation}
    d_\mathrm{com}(z) = c \int_0^z \frac{d\hat{z}}{H(\hat{z})},
	\label{eq:dcom_z}
\end{equation}
where $c$ is the speed of light in vacuum, and $H$ the Hubble parameter, a function of redshift and the cosmological parameters $(H_0, \Omega_m, \Omega_\Lambda)$:
\begin{equation}
    H(z) = H_0 \sqrt{\Omega_m(1+z)^3 + \Omega_\Lambda}.
	\label{eq:H_z}
\end{equation}
Here, $H_0~=~H(0)$ is the Hubble constant, $\Omega_m$ the already mentioned matter fraction of the Universe today, and $\Omega_\Lambda~=~1~-~\Omega_m$ the dark energy fraction today.

As can be seen from the above equations, it is necessary to assume fiducial values for $H_0$ and $\Omega_m$ in order to estimate comoving distances from the observables.
A bad selection of these values will lead to a wrong estimation of $(\sigma, \pi)$, and hence, a distorted spatial distribution of galaxies around voids on comoving coordinates.
This also happens when measuring, for instance, the galaxy auto-correlation function or the power spectrum.
This phenomenon is known as the \citet{ap} effect.

Hereafter, we will refer to this kind of distortions in the spatial distribution of galaxies as \textit{geometrical distortions} (GD).


\subsection{Dynamical distortions}
\label{subsec:rsd}

Besides the GD described in the previous section, there also exist \textit{dynamical} or \textit{redshift space distortions} \citep[RSD]{rsd_kaiser}.
The peculiar velocity of a galaxy along the LOS, $v_\parallel$, generates an additional shift on the spectrum lines due to the Doppler effect, which is indistinguishable from the cosmological redshift due to the universal expansion.
This additional shift affects $z$, and hence, distorts the estimation of the LOS comoving separation $\pi$. 
The apparent comoving separations ($\sigma$, $\pi$) can be written in terms of their respective true comoving separations ($r_\perp$, $r_\parallel$) as follows:
\begin{equation}
    \pi = r_\parallel + \frac{v_\parallel}{H(z)} (1+z) \\
    \sigma = r_\perp.
	\label{eq:distance_rsd}
\end{equation}
Note that $\sigma$ remains unaffected.

Hereafter, we will distinguish between: i) the \textit{observable space} $(\theta, \zeta)$, where measurements are made; ii) the \textit{real space} $(r_\perp, r_\parallel)$, not affected by distortions; and iii) the \textit{distorted space} $(\sigma, \pi)$, where GD and RSD are jointly observed.
Here, $\zeta~:=~|z~-~z'|$.


\subsection{The projected void-galaxy cross-correlation functions}
\label{subsec:tools}

As we mentioned in Section~\ref{sec:intro}, the void-galaxy cross-correlation function is a natural tool to perform an AP test.
According to the cosmological principle, the real-space correlation function, $\xi(r_\perp, r_\parallel)$, possesses spherical symmetry with circular isocontours.
However, when working with observational data, it is only possible to obtain the distorted-space correlation function, $\xi(\sigma, \pi)$, which is not isotropic due to the presence of GD and RSD.
Nevertheless, these anisotropies can be modelled, and this model depends on the cosmological parameters.

Working directly on observable-space, $\xi(\theta, \zeta)$, has the advantage that it is not necessary to assume a fiducial cosmology.
This is the first aspect we incorporated in our method.
The second aspect is related to a third type of systematics besides GD and RSD that can affect the cosmological inference when modelling the correlation function.
Models evaluate the correlation function on a given point of the space.
However, when measuring it, a binning scheme is used, and hence, several scales are mixed.
This is not a problem if we work with almost differential bins, nevertheless, this implies a poor signal.
On the contrary, increasing the bin sizes improves the signal, but the correlation function must then be modelled taking into account the volume and geometry of the bins.
Such a model allows to work with bins of arbitrary sizes, so we can do even more and work with fully projected correlation functions.

If we project $\xi(\theta, \zeta)$ towards the POS in a given redshift range, we get the \textit{plane-of-sky correlation function}, $\xi_\mathrm{pos}(\theta)$, which depends only on the angular coordinate $\theta$.
On the other hand, if we project $\xi(\theta, \zeta)$ towards the LOS in a given angular range, we get the \textit{line-of-sight correlation function}, $\xi_\mathrm{los}(\zeta)$, which depends only on the redshift-difference coordinate $\zeta$.
We measure and model these two complementary functions taking into account the effects of GD, RSD and the scale mixing due to the projection range, and in this way, we perform an AP test to constrain $\Omega_m$ and $\beta$\footnote{$\beta$ will be defined in Section~\ref{subsubsec:model_vel}.}.
We will explain in detail how to define the binning scheme to measure the projected correlations in Section~\ref{subsec:binning}.

Given that GD are sensitive to the redshift of void identification (Eqs.~\ref{eq:distance}), we analyse the performance of our test with the redshift, using void samples up to $z'~=~1.5$. 
This is important in view of the advent of the new generation of galaxy spectroscopic surveys, such as HETDEX \citep{hetdex}, Euclid \citep{euclid}, and DESI \citep{desi}, which in general, will cover a volume with a median redshift larger than $0.5$, a significant improvement with respect to the available surveys.


\section{Data set}
\label{sec:data}

\subsection{Simulation setup}
\label{subsec:simulation}

We used the \textit{Millennium XXL} N-body simulation \citep[MXXL]{mxxl_angulo}.
This simulation extends the previous Millennium and Millennium-II simulations (\citealt{millennium_springel}, \citealt{millennium2_boylan}) and follows the evolution of $6720^3$ dark matter particles inside a cubic box of length $3000~\hmpc$.
The particle mass is $8.46\times10^9~\hmsun$.
This simulation adopts a flat $\Lambda$CDM cosmology with the same cosmological parameters as the previous Millennium counterparts: $\Omega_m~=~0.25$, $\Omega_\Lambda~=~0.75$, $\Omega_b~=~0.045$, $\Omega_\nu~=~0.0$, $h~=~0.73$\footnote{The Hubble constant is usually parametrised as $H_0~=~100~h~\mathrm{Mpc~km^{-1}~s^{-1}}$.
Hence, all distances and masses are expressed in units of $\hmpc$ and $\hmsun$ respectively.}, $n_s~=~1.0$ and $\sigma_8~=~0.9$.
We used three snapshots belonging to redshifts $0.51$, $0.99$ and $1.50$.

We used dark matter haloes as tracers, which were identified as groups of more than $60$ particles using a friends-of-friends algorithm with linking parameter equal to $0.2$ of the mean inter particle separation.
We selected a lower mass cut of $5\times10^{11}~\hmsun$.

Positions and peculiar velocities in real space are available to quantify the effects of distortions.
In order to generate RSD, we picked the z-axis of the simulation box as the LOS direction, and applied Eq.~(\ref{eq:distance_rsd}) to shift all haloes from real to distorted space.
Moreover, the redshift of each snapshot, $z_{\rm box}$, was assumed to be the redshift for void identification.
As these are high redshifts, it is valid to adopt the plane-parallel approximation, where changes in the LOS direction with the observed angles on the sky are neglected.


\subsection{Void catalogues and selected samples}
\label{subsec:voids}

We applied the void finding method described in \citet{clues3}, which is a modified version of the \citet{voids_padilla} algorithm.
Briefly, the algorithm starts with the identification of the largest spherical regions where the overall density contrast satisfies the criterion $\Delta(R_{\rm void})~=~\Delta_{\rm cut}^{\rm id}(z)$, where $R_{\rm void}$ is the radius of the region and $\Delta_{\rm cut}^{\rm id}$ is a redshift dependent threshold obtained from the spherical collapse model \citep{sphcollapse1,sphcollapse2} by fixing a final spherical perturbation of $\Delta_{\rm cut}^{\rm id}(0)~=~-0.9$.
The list of void candidates is then cleaned so that each resulting sphere does not overlap with any other.
Therefore, voids are defined as underdense spherical regions with a well defined centre and radius.

\begin{figure}
\includegraphics[width=\columnwidth]{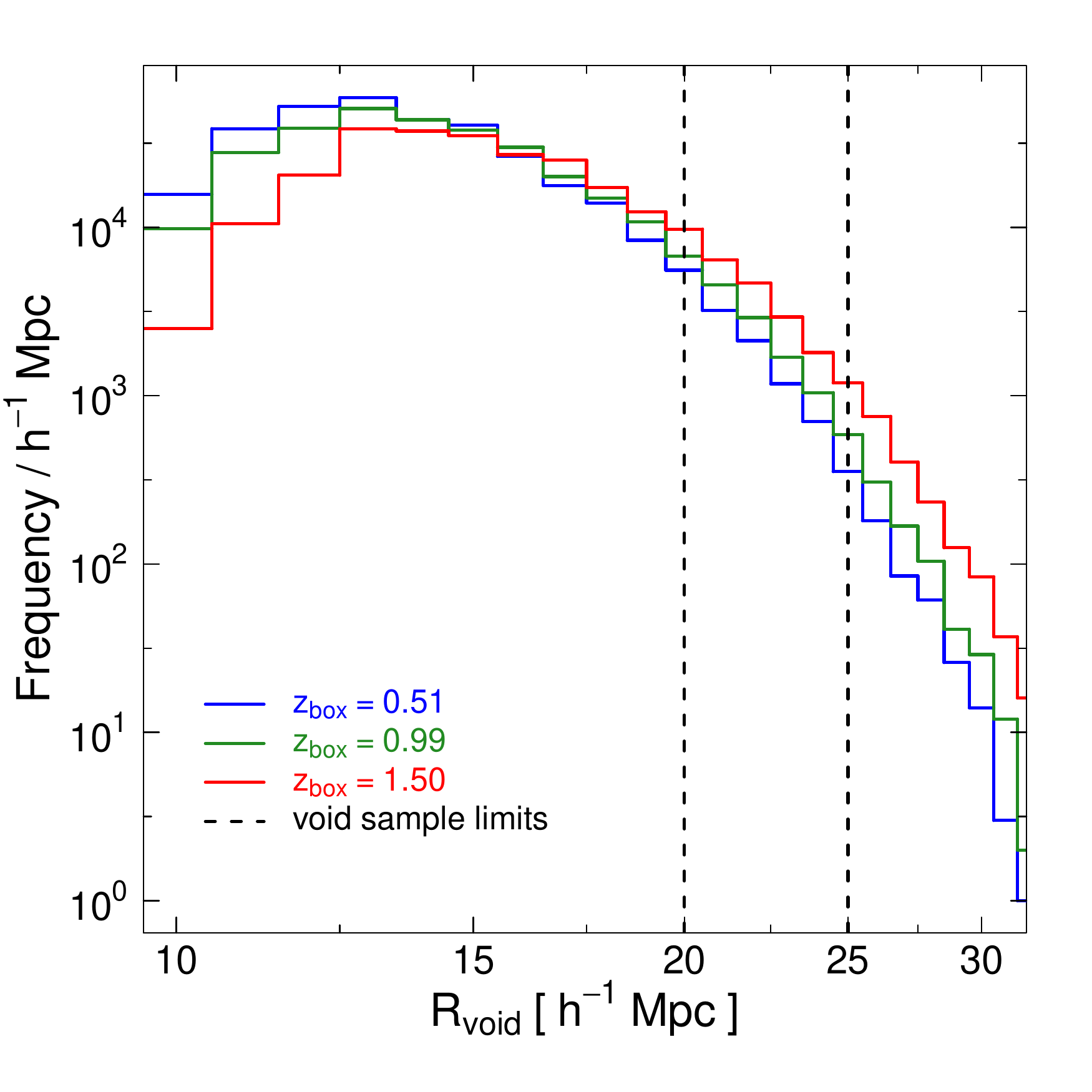}
\caption{
Void radii distribution for each MXXL snapshot.
Vertical dashed lines delimit the void samples used to perform the cosmological test.
}
\label{fig:hist}
\end{figure}

It is worth mentioning that the void finder was applied in real space. Hence, it is cosmology dependent, as well as the values of $\Delta_{\rm cut}^{\rm id}$ and $R_{\rm void}$.
The scope of this paper, however, is to present a non-fiducial test given a galaxy redshift catalogue and a set of underdense centres.
We leave for future investigation the question of void identification in a non-fiducial way, namely, in observable space.
It will also be necessary to compare different void finders and to study the non-trivial effects of finding voids in Mpc-scales.
\citet{voidid_nadathur} analyse the impact of identifying voids in real, redshift and reconstructed real space over the void-galaxy correlation function for the case of the ZOBOV void finder \citep{Zobov}.
In this way, in order to compute distances and densities, needed in void definition, we adopted the same cosmology of the MXXL simulation.
The position of the void centres in comoving coordinates are then transformed into observable-space coordinates with this assumed cosmology.
Table~\ref{tab:catalogues} shows the main characteristics of our halo and void catalogues, and Figure~\ref{fig:hist}, the void radii distribution.

\begin{table}
\centering
\caption{
Main characteristics of the halo and void catalogues.
From left to right: MXXL snapshot, number of dark matter haloes, density threshold criterion for void identification, and number of identified voids.
}
\label{tab:catalogues}
\begin{tabular}{ccccc}
\hline
$z_\mathrm{box}$ & Haloes & $\Delta_{\rm cut}^{\rm id}$ & Voids \\
\hline
\hline
0.51 & 136993439 & -0.8764 & 333741 \\ 
0.99 & 133688808 & -0.8533 & 305082 \\ 
1.50 & 118244901 & -0.8302 & 254993 \\ 
\hline
\end{tabular}
\end{table}

For each MXXL snapshot, we selected voids with sizes between $20~\leq~R_\mathrm{void}/\hmpc~\leq~25$ to perform the cosmological test. 
This is shown with vertical dashed lines in Figure~\ref{fig:hist}.
Our results are not sensitive to the size of the voids used, as we have verified.
However, large void sizes ensure that they are mostly in expansion (see Section~\ref{subsec:densvel}), which simplifies the RSD modelling (see Section~\ref{subsec:model_densvel}).
The number of voids found in each sample are the following: $10157$ for $z_{\rm box}~=~0.51$, $13703$ for $z_{\rm box}~=~0.99$, and $21034$ for $z_{\rm box}~=~1.50$.


\section{Measuring the projected cross-correlation functions}
\label{sec:measurements}

In this section, we explain how to measure the projected POS and LOS correlation functions, $\xi_{\rm pos}(\theta)$ and $\xi_{\rm los}(\zeta)$, and present the results from data.


\subsection{Binning scheme and projection range}
\label{subsec:binning}

As we explained in Section~\ref{subsec:tools}, correlations are isotropic on real space.
However, this is not the case on distorted space due to the presence of the coupled GD and RSD.
If only RSD were present, this spherical symmetry would be reduced to a cylindrical symmetry along the LOS direction.
Actually, as GD are also present, the spherical symmetry is distorted in a more complex way.
Nevertheless, it is still instructive to visualize a void sample as a two dimensional stack with cylindrical axes.

The void-halo cross-correlation function is measured in terms of the void-centric observables $(\theta, \zeta)$, with $\zeta~=~|z~-~z'|$, which is estimated by counting void-halo pairs within a cylindrical binning scheme.
In this geometry, a bin is a cylindrical shell oriented along the LOS, with internal radius $\theta_\mathrm{int}$, external radius $\theta_\mathrm{ext}$, a lower height $\zeta_{\rm low}~=~|z_{\rm low}~-~z'|$, and upper height $\zeta_{\rm up}~=~|z_{\rm up}~-~z'|$.
In order to estimate the correlation value for a given bin, $\xi_{\rm bin}(\theta, \zeta)$, the number of counted pairs, $\mathrm{DD}$, must be normalised by the expected number of pairs in a homogeneous distribution, $\mathrm{DR}$.
In this work, we used the \citet{estimator_davis} estimator:
\begin{equation}
    \xi_{\rm bin}(\theta, \zeta) = \frac{\rm DD}{\rm DR} - 1.
    \label{eq:estimator}
\end{equation}
Here, $(\theta, \zeta)$ denotes the coordinates of the geometrical centres of the bins merely in order to label them.
When working with real data, the \citet{estimator_landy} estimator must also be analysed.

The POS and LOS correlation functions are special cases of this binning scheme.
On the one hand, the scheme for the $\xi_{\rm pos}(\theta)$ correlation consists of a set of nested cylindrical shells across the POS centred at different angles $\theta$, with constant thickness $\delta \theta~:=~\theta_{\rm ext}~-~\theta_{\rm int}$ (the \textit{POS binning step}), and a constant height $\mathrm{PR}_\zeta~:=~\zeta_{\rm up}$ (the \textit{redshift projection range}).
Note that in this case, $\zeta_{\rm low}~=~0$.
On the other hand, the scheme for the $\xi_{\rm los}(\zeta)$ correlation consists of a succession of filled cylinders along the LOS centred at different redshift-separations $\zeta$, with constant length $\delta \zeta~:=~\zeta_{\rm up}~-~\zeta_{\rm low}$ (the \textit{LOS binning step}), and a constant radius $\mathrm{PR}_\theta~:=~\theta_{\rm ext}$ (the \textit{angular projection range}).
Note that in this case, $\theta_{\rm int}~=~0$.

\begin{figure*}
\includegraphics[width=88mm]{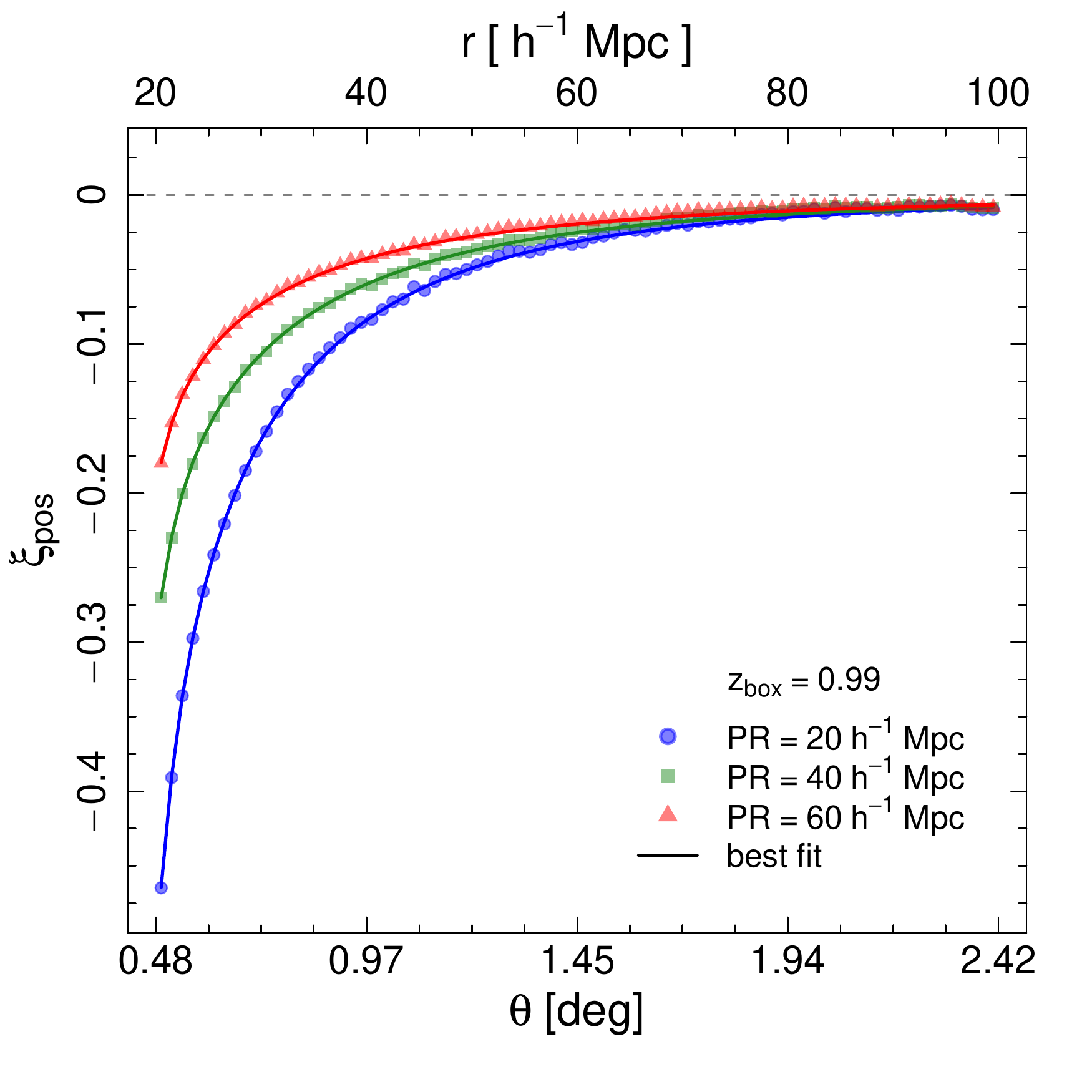}
\includegraphics[width=88mm]{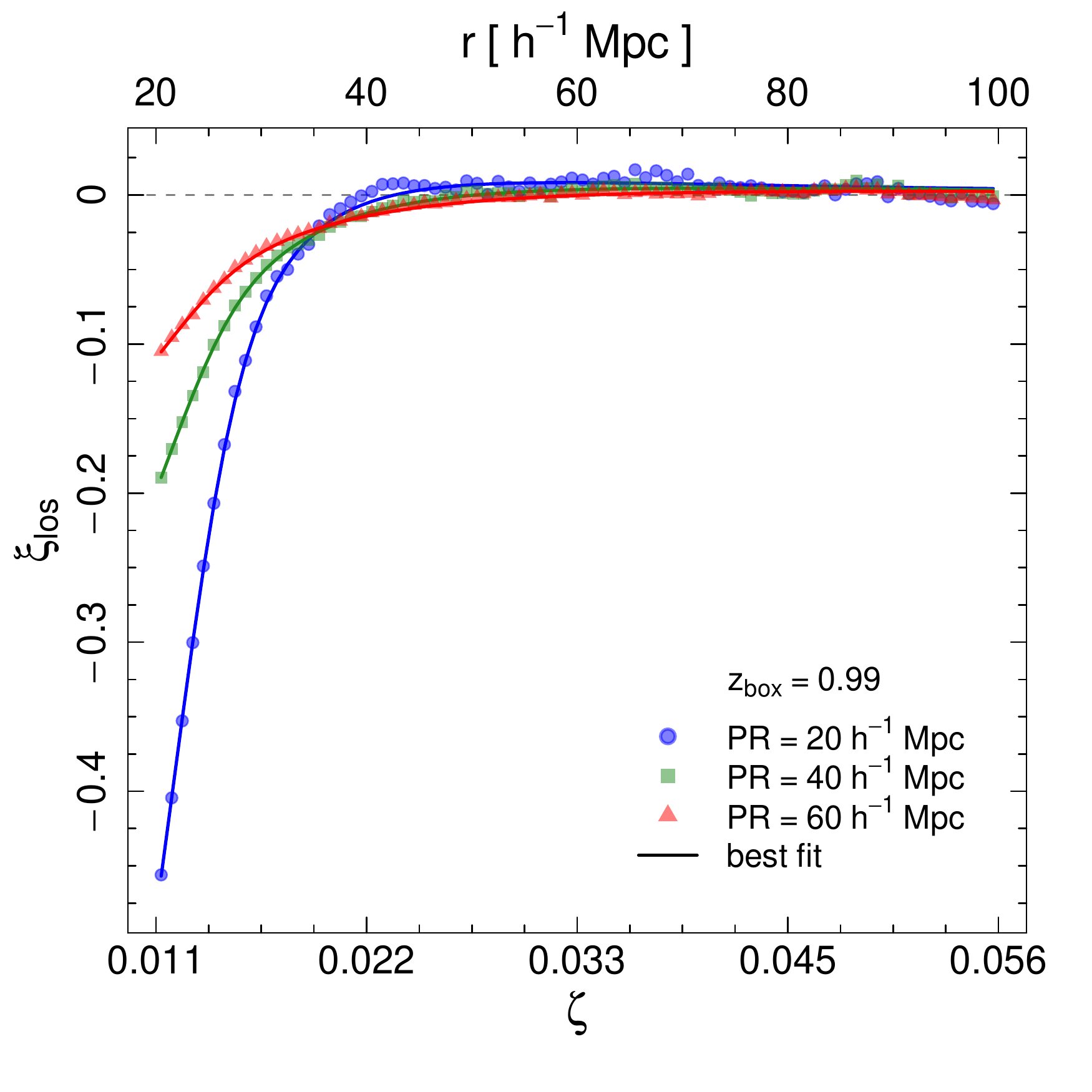}
\caption{
Plane-of-sky (left) and line-of-sight (right) void-halo cross-correlation functions of the void sample taken from the $z_{\rm box}~=~0.99$ MXXL snapshot.
Here are shown the cases ${\rm PR}/\hmpc = 20$ (blue circles), $40$ (green squares) and $60$ (red triangles).
The error bars are not shown because they are smaller than the data points.
Both observable (below) and real-space (above) axes are shown for comparison.
Solid curves are the theoretical functions obtained after the application of the model from Section~\ref{sec:model} with the best fit parameters from Section~\ref{sec:mcmc_fit}.
}    
\label{fig:correlations}
\end{figure*}

For simplicity, we will refer to both $\mathrm{PR}_\zeta$ and $\mathrm{PR}_\theta$ as a single value $\mathrm{PR}$ expressed on real-space units of $\hmpc$ using the MXXL cosmology.
The same applies to the binning steps $\delta \theta$ and $\delta \zeta$.
In this work, we analysed $8$ different ${\rm PRs}/\hmpc$: $1$, $5$, $10$, $20$, $30$, $40$, $50$ and $60$.
Figure~\ref{fig:correlations} shows $\xi_{\rm pos}(\theta)$ and $\xi_{\rm los}(\zeta)$ of the void sample taken from the $z_{\rm box}~=~0.99$ MXXL snapshot for the cases ${\rm PR}/\hmpc~=~20$ (blue circles), $40$ (green squares) and $60$ (red triangles).
The binning step is $\delta \theta~=~\delta \zeta~=~1\hmpc$.
The error bars are not shown because they are smaller than the data points.
The remaining snapshots show a similar behaviour.
By way of comparison, we show both the observable (below) and real-space (above) axes.
We also show the theoretical functions (solid curves) obtained after the application of the model from Section~\ref{sec:model} with the best fit parameters obtained in Section~\ref{sec:mcmc_fit}.
From the figure, it can be seen that the profiles flatten with increasing PRs, and that $\xi_{\rm los}(\zeta)$ is more affected by RSD than $\xi_{\rm pos}(\theta)$.

\subsection{Real-space profiles}
\label{subsec:densvel}

Before we present the physical model for the POS and LOS correlation functions, first it is instructive to study the measured stacked real-space density and velocity profiles of the void samples, as they are the fundamental components of this model (see Section~\ref{subsec:model_densvel}).
These profiles describe the void environment and dynamics.
Although this analysis is not possible with real data as we cannot work on real space, with a simulation instead, we can take advantage of the real-space positions and velocities of tracers.
In this way, we can check if our method is capable of recovering these profiles. This is part of the calibration of our test (more details in Section~\ref{subsec:fit}).


\subsubsection{Density contrast profiles}
\label{subsubsec:density}

\citet{voids_sheth} proposed that the evolution of voids is strongly determined by their surrounding environment.
They provide a theoretical framework based on the excursion set formalism where the void population is twofold according to two evolutionary processes: i) expanding voids embedded in an underdense region compared to the mean density of the Universe (\textit{void-in-void} mode), and ii) collapsing voids surrounded by an overdense shell (\textit{void-in-cloud} mode).
Large voids are typically of the first type, whereas small voids, of the second type.
According to \citet{clues1}, \citet{clues2}, \citet{voids_hamaus} and \citet{clues3}, the density contrast profile characterises the void environment. 
They highlight two kinds of profiles that illustrate two distinct behaviours: i) increasing profiles that tend to the mean density of the Universe at large distances (\textit{R-type}), and ii) profiles with a noticeable maximum and a decline at larger distances towards the mean density (\textit{S-type}).
The key aspect is that R-type voids match with the void-in-void mode, whereas S-type voids, with the void-in-cloud mode.

Consider the stacked \textit{differential density contrast profile} of haloes around voids:
\begin{equation}
	\delta(r) := \frac{n(r) - \bar{n}}{\bar{n}},
    \label{eq:delta_dif}
\end{equation}
where $n(r)$ is the number density of haloes within a void-centric spherical shell of radius $r$ and infinitesimal thickness, and $\bar{n}$ the mean number density of haloes in the simulation box.
In the case of a cross-correlation, $\xi(r)~=~\delta(r)$.
Here, $\xi(r_\perp, r_\parallel)$ has been reduced to a one dimensional profile due to the spherical symmetry, such that $r~=~\sqrt{r_\perp^2~+~r_\parallel^2}$.
Left panel of Figure~\ref{fig:densvel} shows $\xi(r)$ (black circles) of the void sample taken from the $z_{\rm box}~=~0.99$ MXXL snapshot.
The remaining snapshots show a similar behaviour.
Note the three reference dashed lines there: i) the horizontal $\xi~=~-1$ line, which indicates total emptiness, as is the case near the void centres, ii) the horizontal $\xi~=~0$ line, which is the mean value of the Universe, and iii) the vertical $r~=~r_\mathrm{cut}$ line, which indicates the minimum void radius of the sample and can be thought as a representative border between the inner parts of the voids and their environment.

Consider now the stacked \textit{integrated density contrast profile}:
\begin{equation}
	\Delta(r) := \frac{1}{V} \int_V \delta(r) dV = \frac{3}{r^3} \int_0^r \delta(r') r'^2 dr',
    \label{eq:delta_int}
\end{equation}
where $V$ is the volume of the void-centric sphere of radius $r$.
For the second equality, spherical symmetry has been applied.
Left panel of Figure~\ref{fig:densvel} shows $\Delta(r)$ (light-blue squares) of the same void sample ($z_{\rm box}~=~0.99$).
As can be seen, the sample is represented by an increasing stacked profile that tends to the mean density of the Universe at large distances, that is, the sample is mostly composed of R-type voids.

For both density profiles, the solid curves are the theoretical functions obtained after the application of the model from Section~\ref{sec:model} with the best fit parameters from Section~\ref{sec:mcmc_fit}.
We chose R-type voids to perform the cosmological test because they are less sensitive to non-linear effects compared to the S-type, and also because we found a simple parametric model that describes very well their stacked profiles (see Section~\ref{subsec:model_densvel}).


\subsubsection{Velocity profiles}
\label{subsubsec:velocity}

\begin{figure*}
\includegraphics[width=88mm]{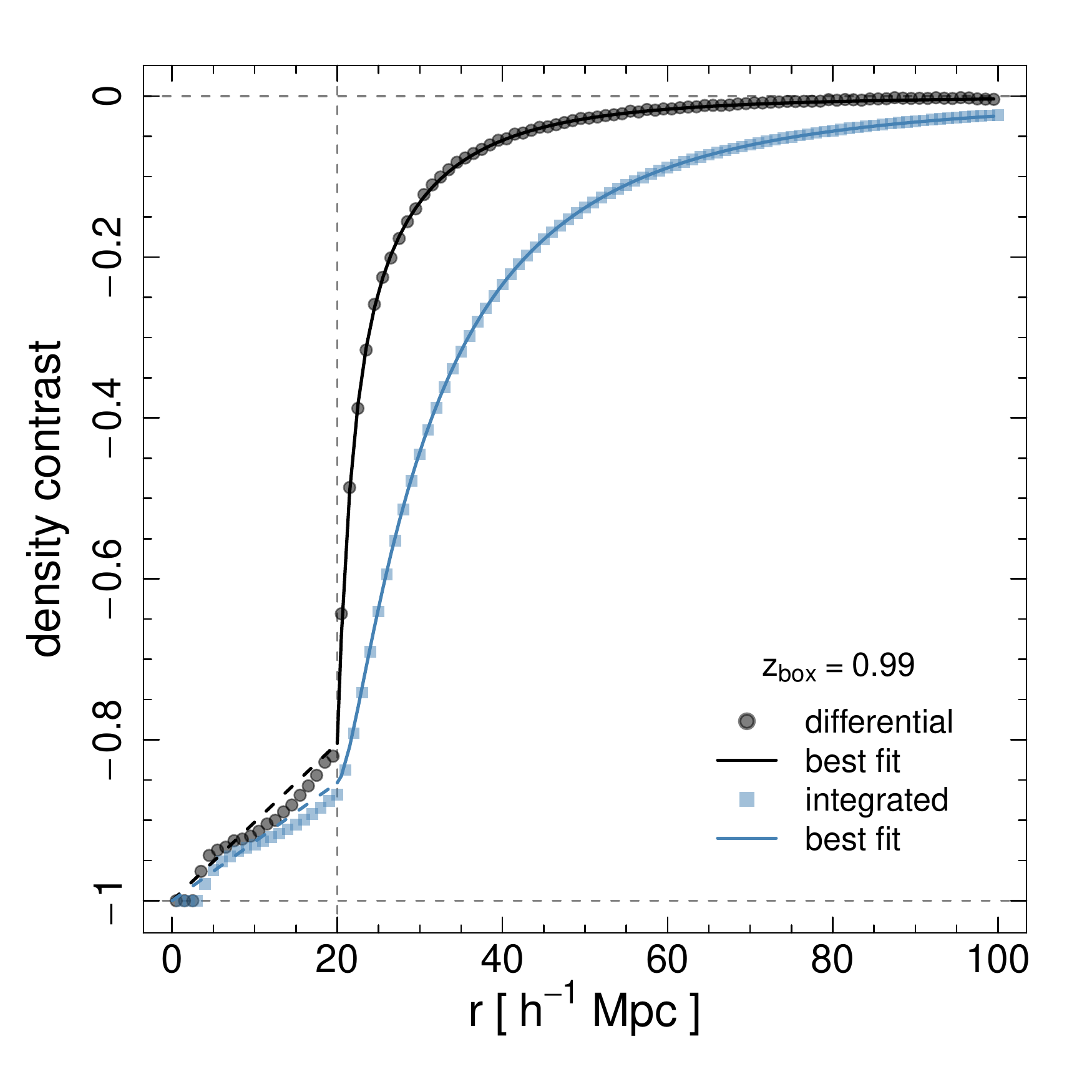}
\includegraphics[width=88mm]{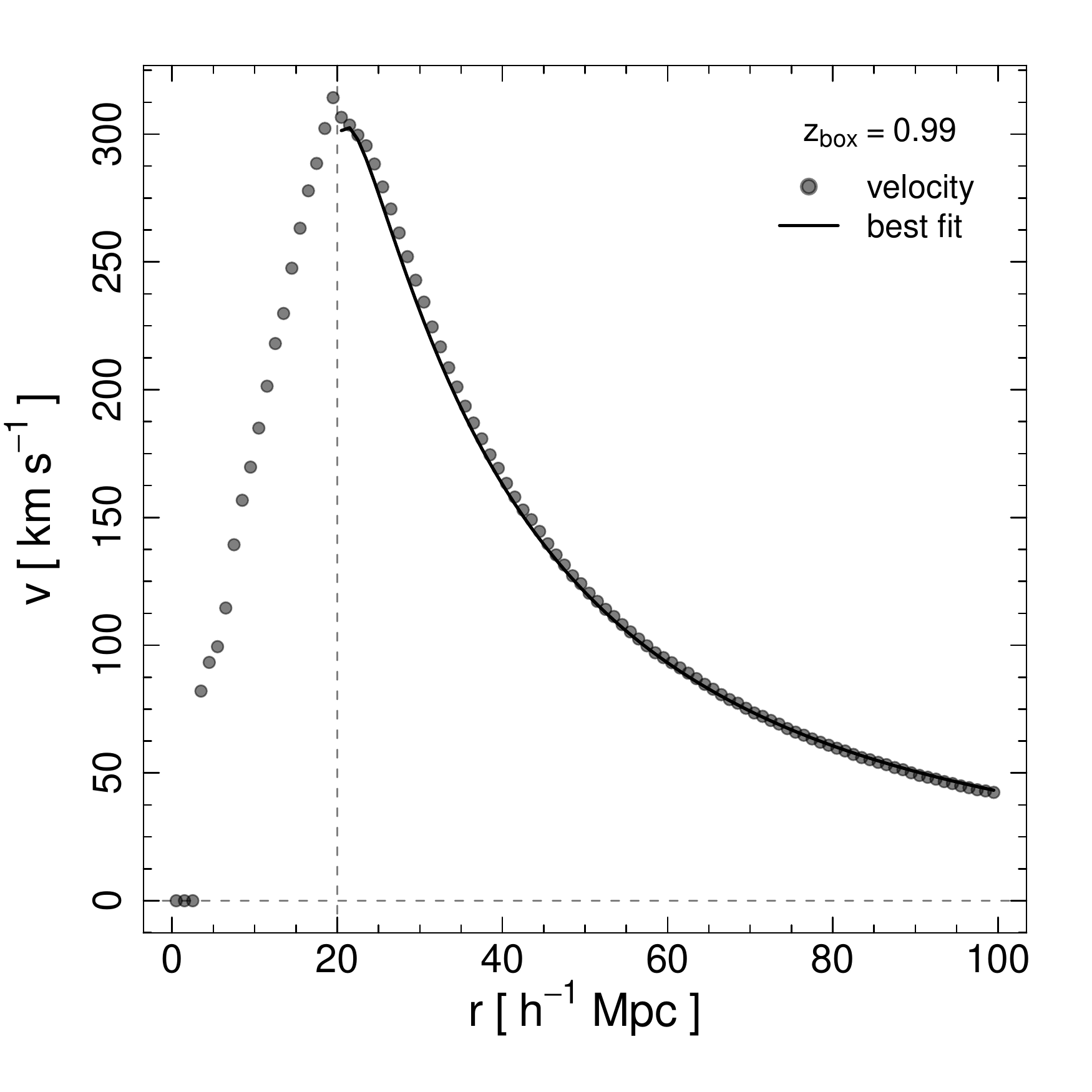}
\caption{
Stacked real-space density and velocity profiles of the void sample taken from the $z_{\rm box}~=~0.99$ MXXL snapshot.
\textit{Left panel.} Differential (black circles) and integrated (light-blue squares) density contrast profiles.
\textit{Right panel.} Radial velocity profile (black circles).
The reference dashed lines are explained in the text.
Solid curves are the theoretical functions obtained after the application of the model from Section~\ref{sec:model} with the best fit parameters from Section~\ref{sec:mcmc_fit} for the case ${\rm PR}=40~\hmpc$.
}
\label{fig:densvel}
\end{figure*}

Right panel of Figure~\ref{fig:densvel} shows the corresponding stacked \textit{radial velocity profile} (black circles). Again, the remaining snapshots show a similar behaviour.
Each point represents the average of the void-centric radial component of the peculiar velocities of haloes around voids.
There are two reference dashed lines: i) the horizontal $v~=~0$ line, which is the mean value of the Universe, and ii) the vertical $r~=~r_\mathrm{cut}$ line. As can be seen, the sample is represented by a profile that shows only expansion velocities, as expected for the R-type voids.
The characteristic $r_\mathrm{cut}$ distance represents the change between increasing (inside the voids) and decreasing (outside the voids) expansion.

\section{Model}
\label{sec:model}

In this section, we present a physical model for the void-halo cross-correlation function on observable space for a general binning scheme $(\theta_\mathrm{int}, \theta_\mathrm{ext}, \zeta_\mathrm{low}, \zeta_\mathrm{up})$.
This model takes into account the effects of GD, RSD and the scale mixing due to the bin sizes.
In this way, the projected POS and LOS correlation functions are treated as special cases with the appropriate bin limits: $(\delta \theta=\theta_{\rm ext}-\theta_{\rm int},0,{\rm PR_\zeta})$ and $(0,{\rm PR_\theta},\delta \zeta=\zeta_{\rm up}-\zeta_{\rm low})$, as was explained in Section~\ref{subsec:binning}.


\subsection{Modelling geometrical distortions and bin size}
\label{subsec:model_gd}

First, we give the full treatment, the one which must be applied when working with observational data.
Then, we turn to the simplifications used for our simulation boxes.


\subsubsection{Full treatment}
\label{subsubsec:model_gd_general}

The first step is to map the distorted-space scales ($\sigma$,$\pi$) that are involved in an observable-space cylindrical bin $(\theta_\mathrm{int}, \theta_\mathrm{ext}, \zeta_\mathrm{low}, \zeta_\mathrm{up})$.
For a fixed void centre at $z'$, there are two $z$ values for each $\zeta$ limit that match the criterion $\zeta~=~|z~-~z'|$.
Specifically, the counted tracers will have angular coordinates between $\theta_\mathrm{int}$ and $\theta_\mathrm{ext}$, and their redshifts $z$ will belong to either one of this two disjoint intervals: $(z'+\zeta_{\rm low}, z'+\zeta_{\rm up})$ or $(z'-\zeta_{\rm up}, z'-\zeta_{\rm low})$.
Therefore, each cylindrical bin corresponds to two volumes on the data. 
Given a set of cosmological parameters, this two regions correspond to two different volumes in distorted space.

Taking this into account, the expected number of data pairs, $\mathrm{DD}$, is given by the following expression\footnote{$\hat{\pi}$ refers to the irrational number \textit{pi}: 3.14159..., to avoid confusion with the $\pi$ coordinate.}:
\begin{equation}
\begin{aligned}
	\mathrm{DD} =& 2\,\hat{\pi}\int_{z'_{\rm min}}^{z'_{\rm max}} dz'~d^2_{\rm com}(z')~n_v(z')~V_{\rm slice}\\
                 &\left[
                 \int_{z'+\zeta_{\rm low}}^{z'+\zeta_{\rm up}}dz~\frac{dd_{\rm com}}{dz}(z)~n_t(z)
                  \int_{\theta_\mathrm{int}}^{\theta_{\rm ext}}d\theta~\theta~\left[1 + \xi(\sigma, \pi)\right]
                  \right.+\\
                 &\left.\int_{z'-\zeta_{\rm up}}^{z'- \zeta_{\rm low}}dz~\frac{dd_{\rm com}}{dz}(z)~n_t(z)
                  \int_{\theta_\mathrm{int}}^{\theta_{\rm ext}}d\theta~\theta~[1 + \xi(\sigma, \pi)]\right],\\
\end{aligned}
    \label{eq:datadata}
\end{equation}
where $n_v(z')$ is the number density distribution of voids in the slice $z'_{\rm min}~\leq~z'~\leq~z'_{\rm max}$ taken from the catalogue to perform the test, $V_{\rm slice}$ the volume of this slice, and $n_t(z)$ the number density distribution of tracers in the bin.
Here, $\xi(\sigma,\pi)$ is a theoretical correlation function defined on distorted space that must be modelled considering the RSD effects.
Its arguments, $\sigma~=~\sigma(\theta,z')$ and $\pi~=~\pi(z,z')$, depend on the observable-space coordinates by means of Eqs.~(\ref{eq:distance}).

The expected number of pairs in a uniform distribution of tracers, $\mathrm{DR}$, is given in a similar fashion:
\begin{equation}
\begin{aligned}
	\mathrm{DR} =& \hat{\pi} \left(\theta^2_{\rm ext}-\theta^2_{\rm int}\right)\int_{z'_{\rm min}}^{z'_{\rm max}} dz'~d^2_{\rm com}(z')~n_v(z')~V_{\rm slice}\\
                 &\left[
                 \int_{z'+\zeta_{\rm low}}^{z'+\zeta_{\rm up}}dz~\frac{dd_{\rm com}}{dz}(z)~n_t(z)
                 + \int_{z'-\zeta_{\rm up}}^{z'- \zeta_{\rm low}}dz~\frac{dd_{\rm com}}{dz}(z)~n_t(z) \right].
\end{aligned}
	\label{eq:datarandom}
\end{equation}
Finally, combining Eqs.~(\ref{eq:datadata}) and (\ref{eq:datarandom}) into Eq.~(\ref{eq:estimator}), we get an estimation of $\xi_{\rm bin}(\theta, \zeta)$.
This is a general expression that takes into account the effect of all possible mixing of scales and GD.


\subsubsection{Simplifications for a simulation box}
\label{subsubsec:model_gd_simu}

The scope of this work is to present a novel cosmological test using voids, focusing our analysis
on the effects of the different types of distortions that arise in the measurements, namely GD, RSD and the scale mixing due to the bin sizes.
In order to test these effects, we used simplified mock catalogues taken from the MXXL as was explained in Section~\ref{subsec:simulation}.
In a forthcoming paper, we will present an analysis on real data, specifically, using the BOSS survey \citep{boss}, for which all the machinery developed above must be applied.

For the case of a simulation box, $n_t$ is a constant function.
We assume a unique redshift for void centres, the one corresponding to the MXXL snapshot, $z_{\rm box}$.
In this way, $n_v(z')$ can be thought as a Dirac-delta distribution.
Therefore, according to Eqs.~(\ref{eq:datadata}) and (\ref{eq:datarandom}), $\xi_{\rm bin}(\theta,\zeta)$ simplifies to the following expression:
\begin{equation}
\begin{aligned}
	\xi_{\rm bin}(\theta, \zeta) = & -1 + 2~\left(\theta^2_{\rm ext}-\theta^2_{\rm int}\right)^{-1}  \\ 
    &\left[\int_{z_{\rm box}+\zeta_{\rm low}}^{z_{\rm box}+\zeta_{\rm up}} \frac{dz}{H(z)}
                  \int_{\theta_\mathrm{int}}^{\theta_{\rm ext}}d\theta~\theta~\left[1 + \xi(\sigma, \pi)\right]\right. +\\
                 &\left.\left.\int_{z_{\rm box}-\zeta_{\rm up}}^{z_{\rm box}- \zeta_{\rm low}}\frac{dz}{H(z)}
                  \int_{\theta_\mathrm{int}}^{\theta_{\rm ext}}d\theta~\theta~[1 + \xi(\sigma, \pi)]\right]\right/\\
                  &\left[\int_{z_{\rm box}+\zeta_{\rm low}}^{z_{\rm box}+\zeta_{\rm up}} \frac{dz}{H(z)} +\int_{z_{\rm box}-\zeta_{\rm up}}^{z_{\rm box}- \zeta_{\rm low}} \frac{dz}{H(z)}~\right],
\end{aligned}
	\label{eq:xi_gd}
\end{equation}
where Eq.~(\ref{eq:dcom_z}) was used to express the $d_{\rm com}(z)$ derivative.

We made a further simplification here, which is to model only the upper $(z_{\rm box}+\zeta_{\rm low}, z_{\rm box}+\zeta_{\rm up})$ integrals and consider the same contribution for the lower ones. This approximation introduces a small difference in the integrated comoving distance spanned by each bin, which 
in all the cases considered is almost negligible. For instance, in the case of $\mathrm{PR_\zeta}~=~20~\hmpc$ at $z_{\rm box}~=~1.50$, this difference translates into an uncertainty of $\sim 0.4\%$, small enough to justify this simplification.

In next section, we give a model for $\xi(\sigma,\pi)$, needed in Eqs.~(\ref{eq:datadata}), (\ref{eq:datarandom}) and (\ref{eq:xi_gd}) that takes into account the RSD effects.


\subsection{Modelling dynamical distortions}
\label{subsec:model_rsd}

Following \citet{xirsdmodel_peebles}, $\xi(\sigma, \pi)$ is computed as the convolution of the real-space correlation, $\xi(r)$, and the pairwise velocity distribution of void-halo pairs, $g(\boldsymbol{r},\boldsymbol{v})$:
\begin{equation}
	1 + \xi(\sigma,\pi) = \int d^3v~[1 + \xi(r)]~g(\boldsymbol{r}, \boldsymbol{v}),
	\label{eq:xi_rsd}
\end{equation}
where $\boldsymbol{r}~=~(r_\perp, r_\parallel)$ and $\boldsymbol{v}~=~(v_\perp, v_\parallel)$ are the respective real-space position and velocity vectors of haloes around voids.
Recall that the real-space coordinates $(r_\perp, r_\parallel)$ are related with their respective distorted-space quantities $(\sigma, \pi)$ by means of Eqs.~(\ref{eq:distance_rsd}).
The pairwise velocity distribution can be chosen as a Maxwell-Boltzmann distribution \citep{clues2, ap1_hamaus, ap2_hamaus}.
Given that only $\pi$ is affected by peculiar velocities, Eq.~(\ref{eq:xi_rsd}) reduces to a 1-dimensional integral via the replacements $g(\boldsymbol{r}, \boldsymbol{v})~\rightarrow~g(r,r_\parallel,v_\parallel)$ and $d^3v~\rightarrow~dv_\parallel$, with $g$ a Gaussian distribution centered on the radial velocity profile $v(r)$, with a constant velocity dispersion, $\sigma_v$:  
\begin{equation}
    1 + \xi(\sigma, \pi) = 
    \int_{-\infty}^{\infty} [1 + \xi(r)] \frac{1}{\sqrt{2\pi}\sigma_v}
    \mathrm{exp} \left[- \frac{(v_\parallel - v(r)\frac{r_\parallel}{r})^2}{2\sigma_v^2} \right]
    \mathrm{d}v_\parallel.
	\label{eq:xi_rsd2}
\end{equation}
In next section, we give a model for $v(r)$ and $\xi(r)$, needed in Eq.(\ref{eq:xi_rsd2}).


\subsection{Modelling the real-space profiles}
\label{subsec:model_densvel}

\subsubsection{Model for $v(r)$}
\label{subsubsec:model_vel}

In the case of voids, density fluctuations around them are moderate compared to the case of virial motions inside clusters, so nonlinearities are expected to be less severe.
The fact that we model a cross-correlation between void centres and haloes, also mitigates nonlinear effects, since the main velocity contribution comes from the tracers and not from the centres.
In contrast, in tracer auto-correlations, the velocities at two locations are correlated with each other, which effectively squares nonlinearities \citep{ap1_hamaus}.
Following linear theory \citep{vel_peebles, clues2, ap1_hamaus, ap2_hamaus}, $v(r)$ can be obtained via mass conservation up to linear order in density:
\begin{equation}
    v(r) = - \frac{1}{3} \frac{H(z)}{(1+z)} \beta(z) r \Delta(r),
	\label{eq:vel}
\end{equation}
where $\beta(z)~=~f(z)/b(z)$ is the ratio between the \textit{logarithmic growth rate of density perturbations}, $f(z)$, and the \textit{linear tracer-mass bias} parameter, $b$.
This bias assumption applies as long as density fluctuations remain moderate: $|\delta(r)|~<~1$ \citep{bias1_pollina,bias2_pollina}.

Within the standard $\Lambda$CDM cosmological model, $f(z)$ can be approximated analytically \citep{fz_linder, ap2_hamaus}:
\begin{equation}
	f(z) \approx \left( \frac{\Omega_m(1+z)^3}{\Omega_m(1+z)^3 + \Omega_\Lambda} \right)^{0.55}.
    \label{eq:f_z}
\end{equation}
As we mentioned in Section~\ref{sec:intro}, theories of modified gravity predict deviations from GR to be most pronounced in unscreened low-density environments, making voids a powerful tool for detecting them.
In such cases, the above equations must be modified.
In order to detect any tension with the standard model, it is instructive to take $f/b$ as a free parameter of the model, and not to incorporate the explicit dependence of $\Omega_m$ on this equation.
In this way, we keep the method general.


\subsubsection{Model for $\xi(r)$}
\label{subsubsec:model_xi}

We have empirically found a parametric model for $\xi(r)$ suitable for our R-type void samples:
\begin{equation}
	\xi(r) =
	\left\{
		\begin{array}{ll}
			Ar - 1 & \mathrm{if} ~ r < r_\mathrm{cut},\\
            -\xi_0 \left[ \left( \frac{r}{r_0} \right)^{-3} + \left( \frac{r}{r_0} \right)^{-\alpha} \right]  &
            \mathrm{if} ~ r \geq r_\mathrm{cut},
		\end{array}
	\right.
	\label{eq:xi_model}
\end{equation}
where $(\xi_0, r_0, \alpha)$ are the three parameters of the model.
Three ranges can be identified: i) the \textit{inner zone} $r~<~r_\mathrm{cut}$, ii) the \textit{environmental zone} $r_\mathrm{cut}~\le~r~\le~100~\hmpc$, and iii) the \textit{outer zone} $r~>~100~\hmpc$.
Remember that $r_\mathrm{cut}$ is the minimum void radius of the sample and represents the border between the inner parts of the voids and their environment (Section~\ref{subsubsec:density}).

Let us start with the environmental zone.
This range is quantified by a double power law with slopes $(-3, \alpha)$ that describes the voids wall and environment.
The other two parameters are an amplitude $\xi_0$, and a pivot distance $r_0$, where the slope changes.
It fails describing the inner zone, since this function tends to $-\infty$ as $r~\rightarrow~0$.
It also fails describing the outer zone beyond $100~\hmpc$ because of the baryonic acoustic oscillation (BAO) feature, a relic clustering excess from the very early Universe.

The inner zone, on the other hand, is not relevant in terms of correlation signal and is not trivial to model.
Moreover, the condition $\Delta(R_{\rm void})~=~\Delta_{\rm cut}^{\rm id}$ imposed by the spherical void finder has a direct effect on the shape of the density and velocity profiles.
This is apparent in Figure~\ref{fig:densvel}, where a discontinuous feature at $r~=~r_{\rm cut}$ can be seen in both panels.
For this reason, we decided to measure correlations only in the environmental zone, as Figure~\ref{fig:correlations} reflects.
Nevertheless, the contribution from the inner and outer zones when reproducing RSD are significant, specially the inner zone.
This is because some scales from the inner zone can be shifted into the environmental zone by means of Eqs.~(\ref{eq:distance_rsd}), making a significant contribution to RSD.
In the same way, some scales from the environmental zone can be shifted into the outer zone. 
Hence, these ranges must be modelled, even though data from there are not used.
For the inner zone, we found that it is sufficient a straight line connecting the points $\xi(0)~=~-1$ and $\xi_{\rm cut}~:=~\xi(r_{\rm cut})$, with the following resulting slope:
\begin{equation}
	A = \frac{1}{r_{\rm cut}}~(\xi_{\rm cut} + 1).
	\label{eq:slope}
\end{equation}
For the outer zone, on the other hand, it is sufficient to extend the scope of validity of the double power law model, finding no significant deviations.

In next section, to finish, we give a derived model for $\Delta(r)$, needed in Eq.(\ref{eq:vel}), and an approximated theoretical value for the slope $A$, since this quantity can only be truly known in real space.


\subsubsection{Model for $\Delta(r)$}
\label{subsubsec:model_deltaint}

The $\Delta(r)$ stacked profile can be modelled combining Eqs.~(\ref{eq:delta_int}) and (\ref{eq:xi_model}):
\begin{equation}
	\Delta(r) =
	\left\{
		\begin{array}{ll}
			\frac{3}{4}Ar - 1 & \mathrm{if} ~ r < r_\mathrm{cut}, \\
            \frac{3}{r^3} \left[ \frac{ar^4}{4} - \frac{r_\mathrm{cut}}{3} + I(r) - I(r_\mathrm{cut}) \right] &
            \mathrm{if} ~ r \geq r_\mathrm{cut}, 
		\end{array}
	\right.
	\label{eq:deltaint_model}
\end{equation}
where $I(r)$ is the indefinite integral of Eq.~(\ref{eq:delta_int}) without the $3/r^3$ pre-factor and with Eq.~(\ref{eq:xi_model}) as integrand:
\begin{equation}
    I(r) = -\xi_0 \left[ r_0^3 \mathrm{ln}(r)
         + \frac{r_0^\alpha}{3-\alpha} r^{3-\alpha} \right].
	\label{eq:integral}
\end{equation}

To give an approximate value for the slope $A$, we realised that the true value $\Delta_{\rm cut}~:=~\Delta(r_{\rm cut})$ can be approximated by the identification-method value $\Delta\mathrm{_{cut}^{id}}$ shown in Table~\ref{tab:catalogues}.
Then, from Eq.~(\ref{eq:deltaint_model}) for $r~<~r_{\rm cut}$,
\begin{equation}
    A \approx \frac{4}{3 r_\mathrm{cut}} (\Delta\mathrm{_{cut}^{id}} + 1).
	\label{eq:slope2}
\end{equation}
From Eqs.~(\ref{eq:slope}) and (\ref{eq:slope2}), we can also give an approximate value for $\xi_{\rm cut}$:
\begin{equation}
    \xi_{\rm cut} \approx \frac{4}{3}(\Delta\mathrm{_{cut}^{id}} + 1) - 1.
	\label{eq:xicut_value}
\end{equation}
The validity of these approximations can be corroborated visually in Figure~\ref{fig:densvel}, where the approximated values for for $\Delta_{\rm cut}$ and $\xi_{\rm cut}$ (where the dashed and solid curves match) are near the corresponding true values (data points at the vertical $r~=~r_{\rm cut}$ reference line).


\section{Testing the method}
\label{sec:mcmc_fit}

In this section, we test how well does our model reproduce the features of the projected POS and LOS correlation functions presented in Section~\ref{sec:measurements}, as well as the corresponding real-space profiles.
The aim is to extract cosmological information from the parameters involved in the model.
These parameters can be summarised in two sets: i) the \textit{cosmological set} $\lbrace \Omega_m, H_0, \beta \rbrace$, and ii) the \textit{nuisance set} $\lbrace \xi_0, r_0, \alpha, \sigma_v \rbrace$.
In order to constrain these parameters, we implemented a likelihood exploration with a \textit{Markov Chain Monte Carlo} (MCMC) technique, using a \textit{Metropolis-Hastings} sampler \citep{metropolis_mcmc_1953,hastings_mcmc_1970}.
We will focus our analysis on the cosmological set, specifically, on $\Omega_m$ and $\beta$.
In this work, we considered a fixed $H_0$ value, the one corresponding to the MXXL (Section~\ref{subsec:simulation}).
When applied to a real data set, $H_0$ can be extracted from a different and independent method.


\subsection{Likelihood analysis}
\label{subsec:mcmc}

Let us denote the \textit{likelihood function} with $\mathcal{L}(\boldsymbol{\theta} | \boldsymbol{x})$, where $\boldsymbol{\theta}~=~\lbrace \xi_0,r_0,\alpha,\sigma_v,\Omega_m,\beta \rbrace$ denotes the parameter space, and $\boldsymbol{x}$ the measured data.
The MCMC chains explores $\mathcal{L}(\boldsymbol{\theta} | \boldsymbol{x})$ near its maximum until they reach the equilibrium distribution, that is, the method finds the $\boldsymbol{\theta}$ values that make the data most probable and their confidence regions.
We took the \citet{gelman_rubin} convergence criterion, which compares the spread of the means between chains with the variance of the target distribution.
Once the chains satisfy this criterion, the unburned parts are discarded and the remaining ones are used to sample the likelihood function.

The $\mathcal{L}$ function is obtained by computing the differences between the measured and modelled correlation functions for a given set of parameters:
\begin{equation}
    \mathrm{ln}(\mathcal{L}) = - \boldsymbol{\Delta\xi}^T \boldsymbol{C}^{-1} \boldsymbol{\Delta\xi} + \mathrm{constant}.
	\label{eq:likelihood}
\end{equation}
Here, both measured and modelled correlations are denoted as single vectors $\boldsymbol{\xi}~:=~(\xi_{\rm pos},\xi_{\rm los})$ containing the correlation values in each bin, and $\boldsymbol{\Delta\xi}$ denotes the corresponding difference vector.
$\boldsymbol{C}$ denotes the associated \textit{covariance matrix}.
Each element $C_{ij}$ is computed on the data by a \textit{jackknife re-sampling} using the multivariate generalization of \citet{efron_jackknife_1982}:
\begin{equation}
    C_{ij} = \frac{n-1}{n} \sum_{k=1}^n \left[ \xi_{(k)} - \xi_{(.)} \right]_i \left[ \xi_{(k)} - \xi_{(.)} \right]_j,
	\label{eq:cov}
\end{equation}
where $n$ is the number of jackknife realisations, $\xi_{(k)}$ the correlation function for the $k^{\rm th}$ jackknife realisation, and $\xi_{(.)}$ the average of $\xi_{(k)}$ over the $n$ realisations.
If $2m$, the number of bins, is the dimension of $\boldsymbol{\xi}$, then $2m~\times~2m$ is the dimension of $\boldsymbol{C}$.
This is by far much smaller than those in the traditional case, where the correlation is a $\boldsymbol{\xi}[m \times m]$ matrix and the covariance is a $\boldsymbol{C}[m^2 \times m^2]$ matrix.
This is a key aspect of our method, first because the estimation of the inverse of a smaller matrix is numerically more stable, and second and more important, because the propagation of covariance errors into the likelihood estimates are substantially reduced, allowing to use a smaller number of mock catalogues \citep{{Taylor13,Dodelson13}}.
In this work, we used $\boldsymbol{\xi}[2*80]$ correlation vectors and $\boldsymbol{C}[160 \times 160]$ covariance matrices.

\begin{figure*}
\includegraphics[width=\textwidth]{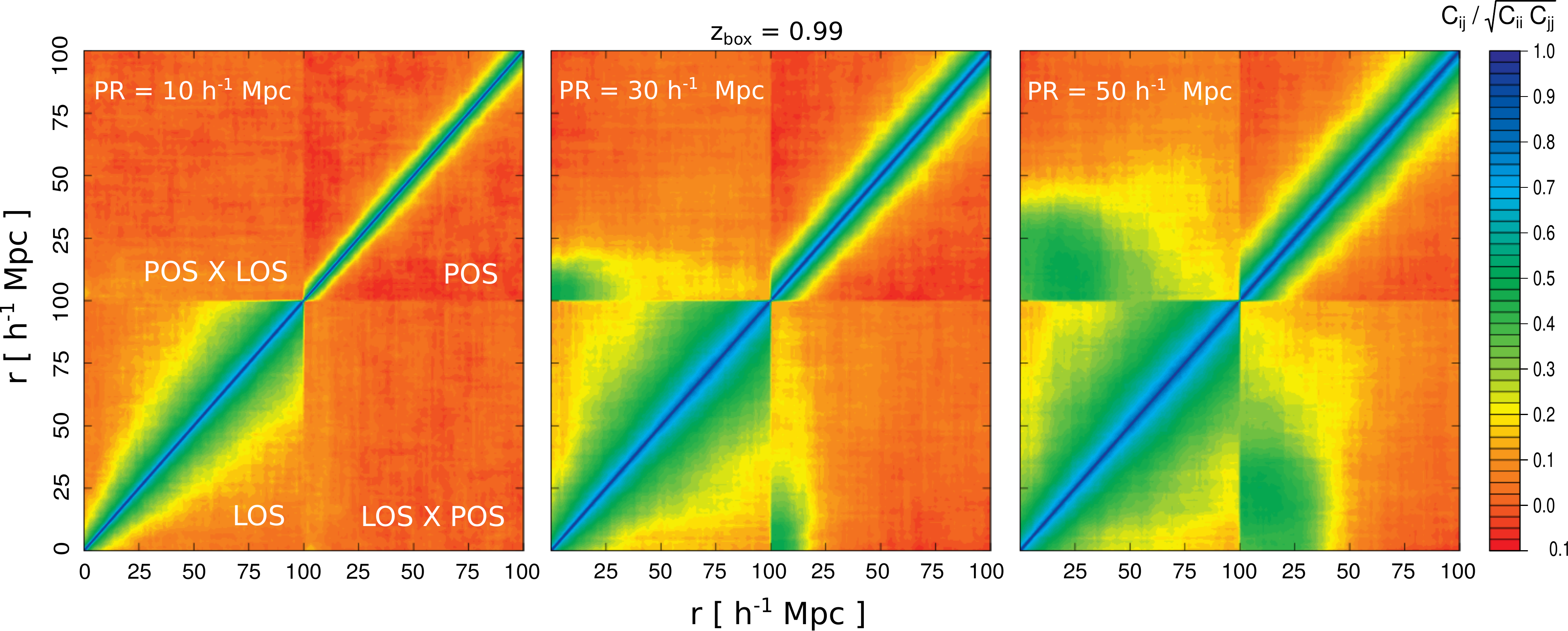}
\caption{
Normalised covariance matrices (correlation matrices), shown as coloured contour maps, of the void sample taken from the $z_{\rm box}~=~0.99$ MXXL snapshot for the cases ${\rm PR}/\hmpc~=~10$, $30$ and $50$.
Absolute values range from $0$ (red) to $1$ (blue).
The x and y-axes are expressed in terms of the real-space coordinates.
Four distinct quadrants can be distinguished on each matrix: the $\xi_{\rm los}$ covariance (bottom left), the $\xi_{\rm pos}$ covariance (top right) and the symmetric $\xi_{\rm los} \times \xi_{\rm pos}$ covariance (bottom right and top left).
The diagonal is the variance.
Clear off-diagonal patterns arise as the PR increases.
}
\label{fig:cov}
\end{figure*}

Figure~\ref{fig:cov} shows the covariance matrices of the void sample taken from the $z_{\rm box}~=~0.99$ MXXL snapshot for the cases ${\rm PR}/\hmpc~=~10$, $30$ and $50$.
The remaining snapshots show a similar behaviour.
Technically, the \textit{correlation matrices} are shown: $C_{ij}/\sqrt{C_{ii}C_{jj}}$, which acquire absolute values from $0$ to $1$, encoded as a coloured contour map from red to blue.
The x- and y-axes are expressed in real-space coordinates for a better comparison.
Note that the matrices are not diagonal since the independence of the correlation values for bins at different scales can not be guaranteed.
In fact, they show clear patterns.
Focusing on an individual matrix, four distinct quadrants can be seen: i) the bottom left quadrant is the covariance sub-matrix for the $\xi_{\rm los}$ correlation, ii) the top right, the covariance sub-matrix for the $\xi_{\rm pos}$ correlation, whereas iii) and iv) the bottom right and top left are the symmetric $\xi_{\rm los}~\times~\xi_{\rm pos}$ covariances.
The diagonal of the entire matrix is the global variance.
The square root of these diagonals make up the implicit error bars in Figure~\ref{fig:correlations}.
Comparing now the three matrices, it can be seen that if the PR is small, the covariance matrices tend to be diagonal, whereas as the PR increases, off-diagonal values become more prominent.
For instance, note the increment of the covariance values on the $\xi_{\rm los}~\times~\xi_{\rm pos}$ quadrants.

\subsection{Cosmological constraints}
\label{subsec:fit}

We present now the results of the likelihood analysis.
Since $\Omega_m$ and $\beta$ are the two parameters of interest, we will focus mainly on them.
The goal is to calibrate the method.
The test is calibrated if we recover the MXXL values of the model parameters.
On the one hand, $\Omega_m^{\rm box}~=~0.25$ (Section~\ref{subsec:simulation}).
On the other hand, the target $\beta_{\rm box}$ values were inferred by fitting directly the real-space velocity profiles measured in the simulation boxes (Section~\ref{subsubsec:velocity}) with Eq.~(\ref{eq:vel}).

\begin{figure*}
\includegraphics[width=88mm]{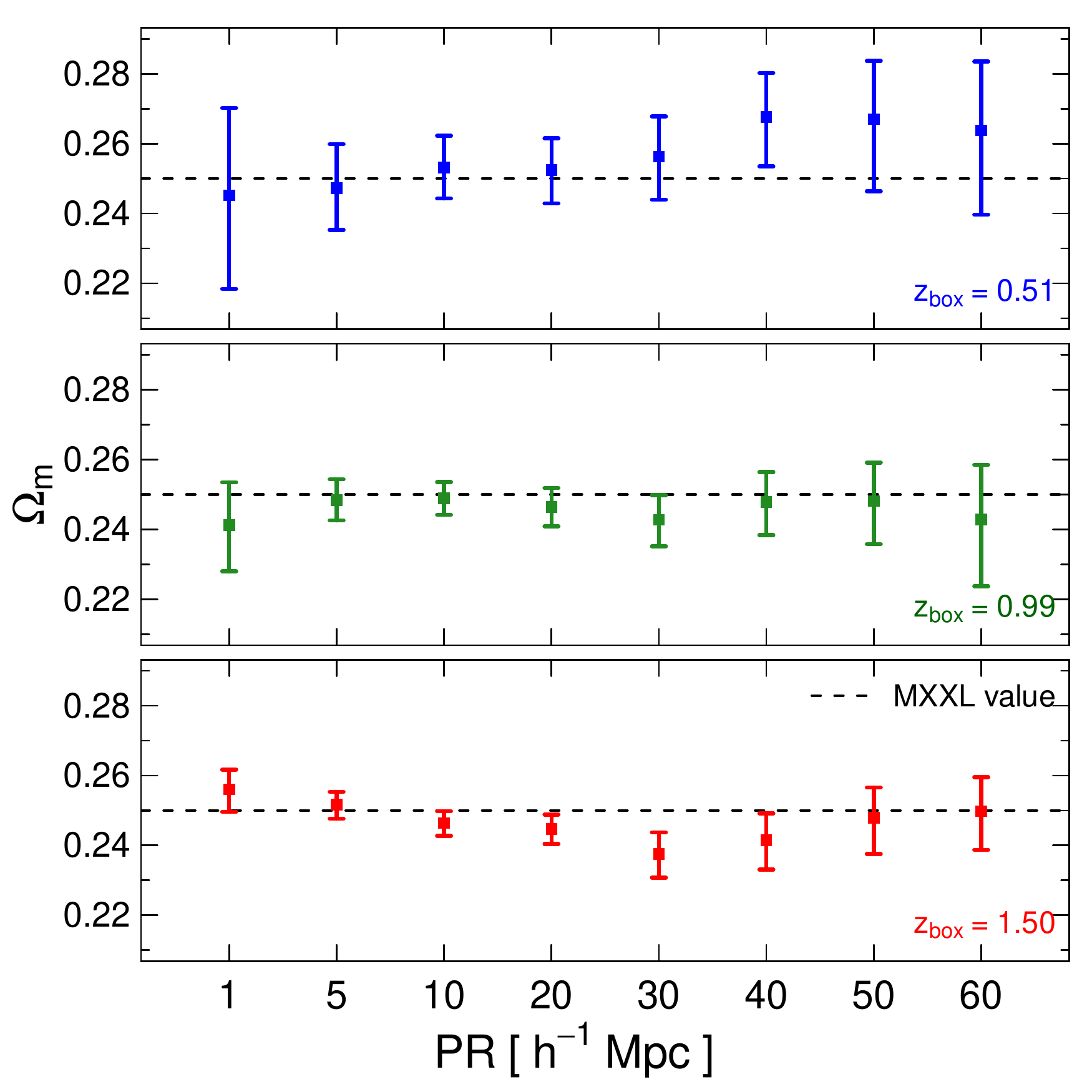}
\includegraphics[width=88mm]{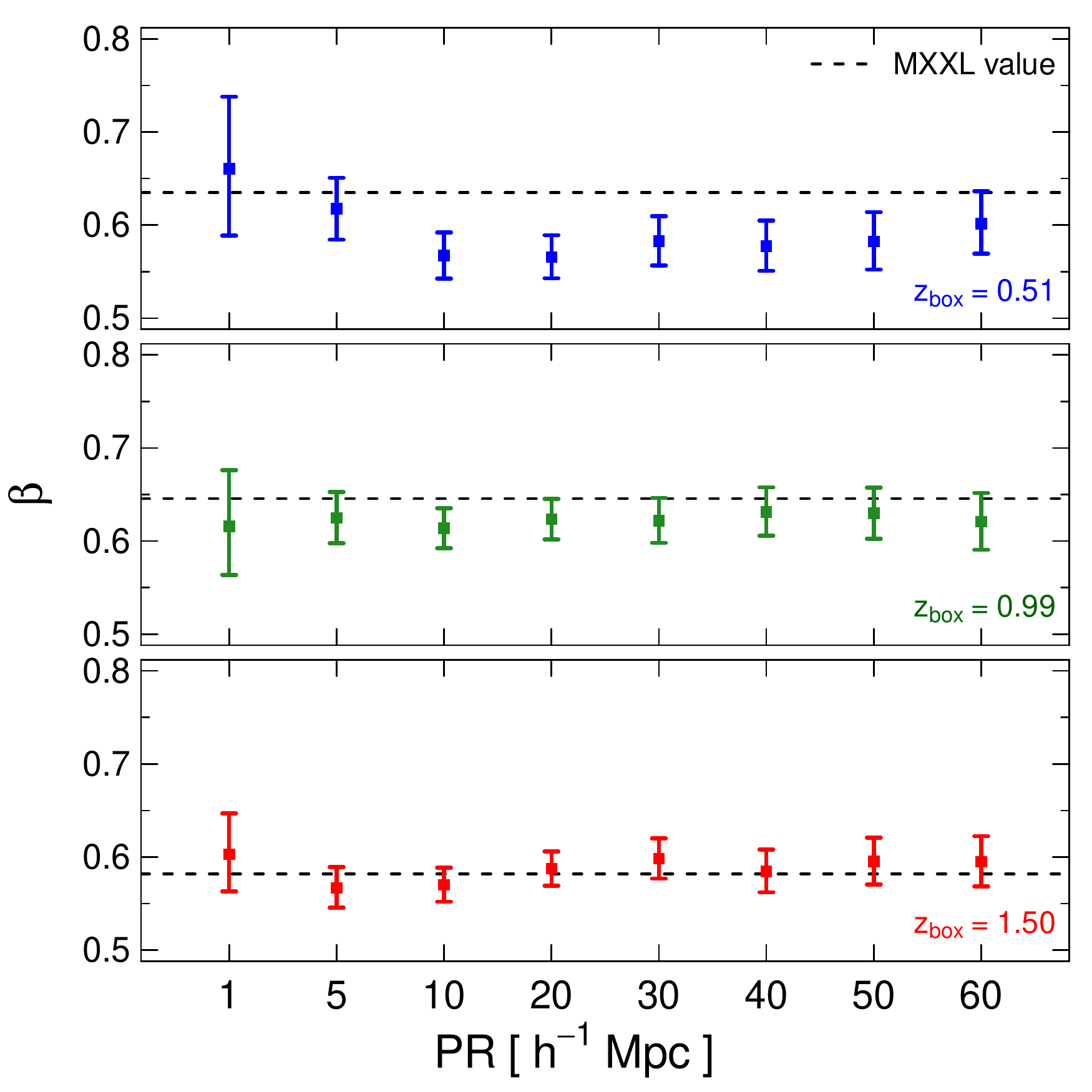}
\caption{
Marginalised likelihood distributions of $\Omega_m$ (left panel) and $\beta$ (right panel) for each MXXL snapshot and for each projection range, shown as $1\sigma$ ($68.3\%$) error bars.
Dashed horizontal lines indicate the MXXL values.
}
\label{fig:boxplots}
\end{figure*}

Figure~\ref{fig:boxplots} shows the $\Omega_m$ and $\beta$ marginalised likelihood distributions for each MXXL snapshot and for each projection range as $1\sigma$ ($68.3\%$) error bars.
These distributions show a Gaussian shape.
The dashed horizontal lines indicate the MXXL values.
Note that $\beta_{\rm box}$ is slightly dependent on $z_{\rm box}$.
As can be seen, the MXXL values fall inside the error bars in most cases, and fall inside $3\sigma$ ($99.7\%$) in all of them, which is the consistency check we were looking for.
This is a consequence of the ability of the model to reproduce GD, RSD and the scale mixing, as can be also corroborated by inspecting Figures~\ref{fig:correlations} and \ref{fig:densvel}, where the theoretical profiles (solid curves) obtained with the best fit parameter values from this likelihood analysis match very well the data points.
However, there is an appreciable deviation with respect to the MXXL value in the case of $\beta$ at $z_{\rm box}~=~0.51$ for ${\rm PR}~\geq~10~\hmpc$.
This is possibly due to a deficiency in the linear model for RSD (Section~\ref{subsec:model_rsd}), as RSD prevail over GD at lower redshifts.
\citet{rsd_achitouv1} and \citet{rsd_nadathur} present an improved RSD model for voids, analysing non-linearities and second order effects.
Note also that the error bars for $\beta$ are almost constant, nearly independent of the PR and $z_{\rm box}$.
In the case of $\Omega_m$, the error bars reach a minimum at PRs between $10$ and $20~\hmpc$, which points out the optimal range to perform the test.
Moreover, they generally decrease from lower to higher $z_{\rm box}$, which elucidates that better confidence regions are obtained performing the test at higher redshifts.
This is due to the fact that GD are more sensitive in the model at higher redshifts.

Two aspects worth mentioning.
First, we are using a high density halo sample in a large volume.
Therefore, the confidence levels on the estimated parameters must be understood in a precision limit framework.
When applied to real data, the confidence regions will be larger.
Second, for a fixed $z_{\rm box}$, the box-plots are not independent estimates, since it is about measuring the same correlation function for the same void sample, merely adding more void-halo pairs when increasing the projection range.
Having said this, Figure~\ref{fig:boxplots} is a robust confirmation that the test can be applied with a wide variety of projection ranges.

\begin{figure*}
\includegraphics[width=\textwidth]{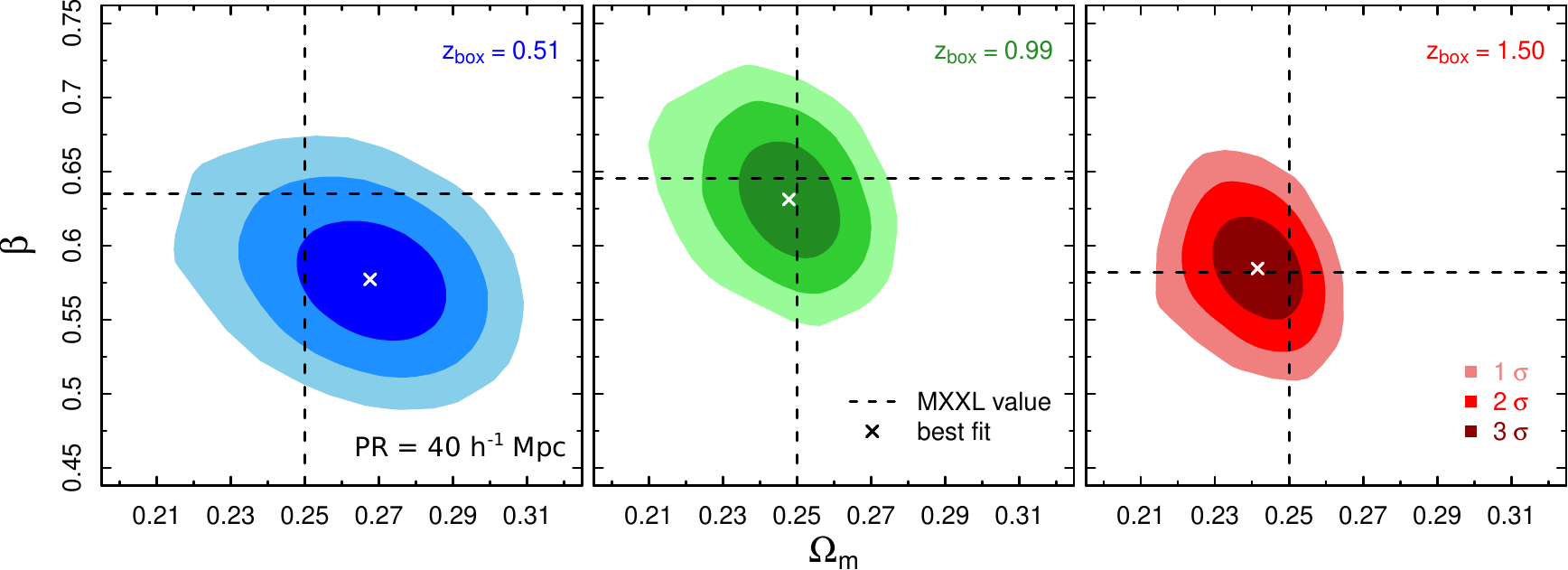}
\caption{
Marginalised likelihood distributions onto the $\Omega_m-\beta$ plane for the case ${\rm PR}~=~40~\hmpc$ for each MXXL snapshot.
From the inner to the outermost, the coloured contour levels enclose $1\sigma$ ($68.3\%$), $2\sigma$ ($95.5\%$) and $3\sigma$ ($99.7\%$) confidence regions.
Dashed lines indicate the respective MXXL values, whereas the white crosses, the best fit values.
}
\label{fig:bananas}
\end{figure*}

\begin{figure*}
\includegraphics[width=\textwidth]{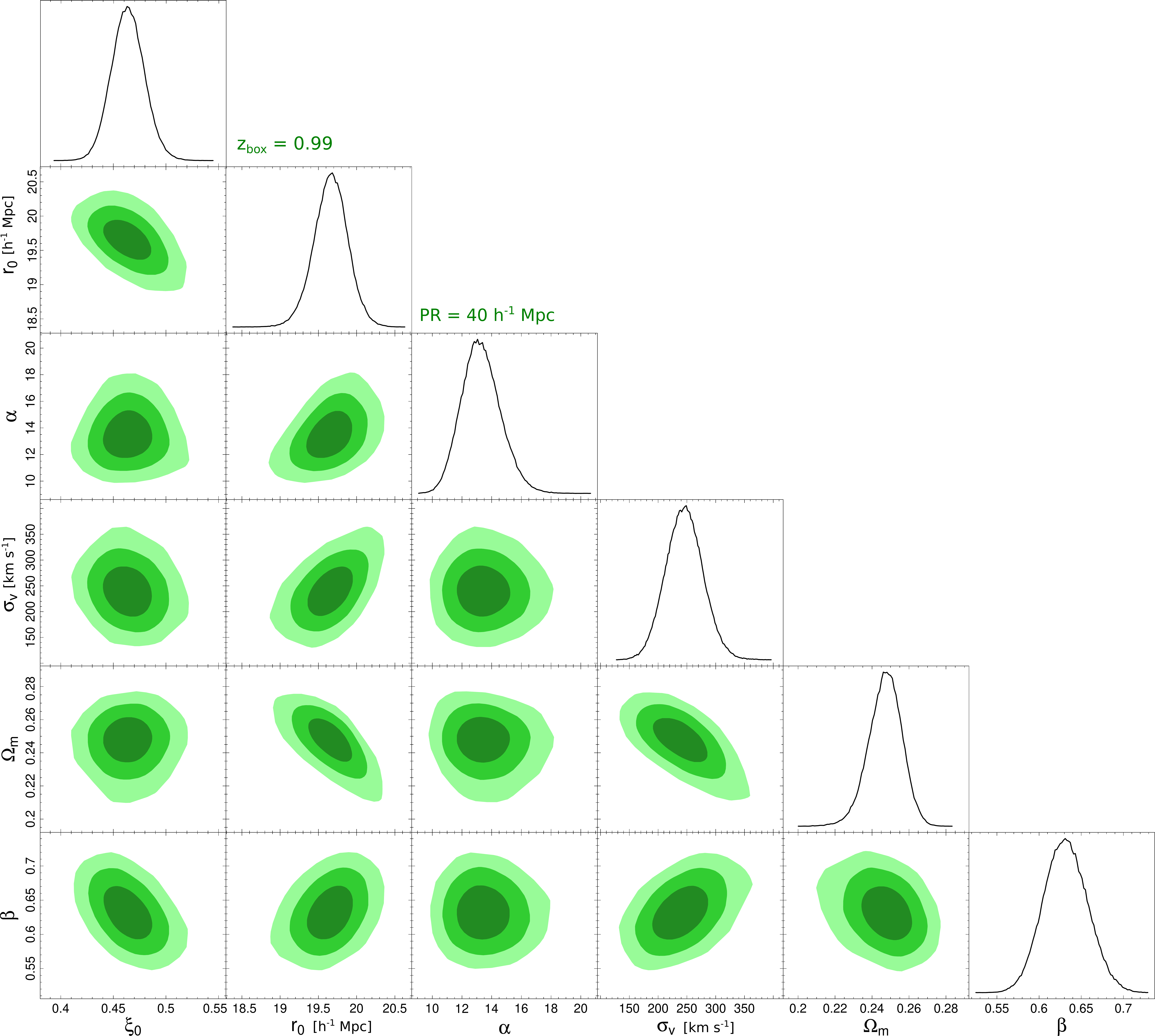}
\caption{
Two dimensional likelihood marginalisations of the full parameter space for the case $z_{\rm box}=0.99$ and ${\rm PR} = 40\hmpc$.
The $\Omega_m-\beta$ panel is the same as the mid panel of Figure~\ref{fig:bananas}.
}
\label{fig:bananas2}
\end{figure*}

Figure~\ref{fig:bananas} shows, as an example, the two-dimensional likelihood marginalisations towards the $\Omega_m-\beta$ plane for the case ${\rm PR}~=~40~\hmpc$ for the three MXXL snapshots.
From the inner to the outermost, the coloured contour levels enclose $1\sigma$ ($68.3\%$), $2\sigma$ ($95.5\%$) and $3\sigma$ ($99.7\%$) confidence regions.
Dashed lines indicate the respective MXXL values, whereas the white crosses, the best fit values.
Note that the target values fall inside the $1\sigma$ confidence region for medium and high redshifts, whereas for low redshift the deviation of $\beta$ explained before can be appreciated.
While the best results (tightest constraint and smallest deviation) were obtained for the ${\rm PR}~=~5~\hmpc$ case, we decided to show the ${\rm PR}~=~40~\hmpc$ case to highlight the robustness of the test with the PR, which is important because the wider the PR is, more data pairs are counted, and therefore, the measured signal increases.
In this sense, this is a a more realistic case applicable to real data.
For completeness, Figure~\ref{fig:bananas2} shows the two-dimensional marginalisations of the full parameter space for the case $z_{\rm box} = 0.99$ and ${\rm PR} = 40\hmpc$.
The constraints are tight, showing no degeneracies among each other.
Moreover, the distributions show a Gaussian shape.
The remaining snapshots and {\rm PRs} show a similar behaviour.

\section{Summary and conclusions}
\label{sec:conclusions}

Cosmic voids are powerful cosmological laboratories.
Their potential will be greatly exploited with the advent of modern galaxy redshift surveys, which will have a median redshift larger then $0.5$, a significant improvement with respect to the available surveys.
The void-galaxy cross-correlation function is a statistical tool that describes the void environment and dynamics.
Taking advantage of the ability to model the coupled dynamical and geometrical distortions which affect it, cosmological information can be extracted, since this model depends on the cosmological parameters.
Dynamical distortions (RSD) arise due to the LOS component of the peculiar velocities of galaxies surrounding voids, whereas geometrical distortions (GD) arise when a wrong cosmology is used to assign a distance scale to measure correlations, the Alcock-Paczynski effect.

There is, in addition, a third type of systematics that affects the cosmological inference when modelling the correlation function.
In the measuring process, a binning scheme is used, and hence, several scales are mixed in the observation.
Increasing the bin sizes improves the signal, but then, models must carefully take into account the volume and geometry of the bins.
Such a model allows to work with bins of arbitrary sizes, so we can do even more and work with fully projected correlation functions.
Moreover, we treat correlations directly in terms of void-centric angular distances and redshift differences $(\theta, \zeta)$ between void-galaxy pairs, so that it is not necessary to assume a fiducial cosmology.
Projecting $\xi(\theta, \zeta)$ towards the POS in a given redshift range, we get the POS correlation function, $\xi_\mathrm{pos}(\theta)$, which depends only on the angular coordinate $\theta$, whereas projecting $\xi(\theta, \zeta)$ towards the LOS in a given angular range, we get the LOS correlation function, $\xi_\mathrm{los}(\zeta)$, which depends only on the void-centric redshift coordinate $\zeta$.
Both projections constitute the fundamental observables of our cosmological test (Figure~\ref{fig:correlations}).

In Section~\ref{sec:model}, we presented a physical model for the void-halo cross-correlation function on observable space for a general cylindrical binning scheme $(\theta_\mathrm{int}, \theta_\mathrm{ext}, \zeta_\mathrm{low}, \zeta_\mathrm{up})$.
This model takes into account the effects of GD, RSD and the mixing of scales due to the bin sizes. The projected POS and LOS correlation functions constitute special cases with the appropriate bin limits.

The goal of our method is to extract cosmological information from the parameters involved in the model.
These parameters can be summarised in two sets: i) the \textit{cosmological set} $\lbrace \Omega_m, \beta \rbrace$, and ii) the \textit{nuisance set} $\lbrace \xi_0, r_0, \alpha, \sigma_v \rbrace$.
$\Omega_m$ is sensitive to GD, whereas $\beta$ is sensitive to RSD.
The method was calibrated using an N-body simulation, the Millennium XXL, for three snapshots: $z_{\rm box}~=~0.51$, $0.99$ and $1.50$.
In order to constrain the parameters, we implemented a likelihood exploration with a MCMC technique. Considering the full set of parameters, the constraints are tight, with no degeneracies among each other, and they show Gaussian distributions.
The main results are presented in Figure~\ref{fig:boxplots}, which shows the likelihood marginalisations of $\Omega_m$ and $\beta$ as $1\sigma$ ($68.3\%$) error bars.
The MXXL values fall inside them in almost all cases, which is a consistency check of the reliability of the method.
It also shows robustness with the projection range.
This robustness is strengthened, in addition, by the fact that the theoretical curves obtained after the application of the model to the best fit parameters from the likelihood analysis match the data points of the measured POS and LOS correlations (Figure~\ref{fig:correlations}), as well as the density and velocity profiles (Figure~\ref{fig:densvel}).
The appreciable deviation of $\beta$ with respect to the MXXL value at $z_{\rm box}~=~0.51$ for ${\rm PR}~\geq~10~\hmpc$, can be attributed to a possible deficiency in the linear model for RSD (Section~\ref{subsec:model_rsd}), as RSD prevail over GD at lower redshifts.
The error bars in the case of $\Omega_m$, show that there is an optimum {\rm PR} and that better confidence regions are obtained performing the test at higher redshifts, where GD are more sensitive in the model.

It is worth mentioning that the method presented here is a non-fiducial test given a galaxy spectroscopic catalogue and a set of underdense centres.
The void identification is a difficult task that deserves particular attention before applying the test to observational data.
A non-fiducial way of finding voids has still to be found and the non-trivial effects of identification in Mpc-scales must be completely understood (Section~\ref{subsec:voids}), a topic for future investigation.

Finally, the data covariance matrices (Figure~\ref{fig:cov}) associated to the projected POS and LOS correlation functions are much smaller than the traditional ones.
Therefore, the propagation of covariance errors into the likelihood estimates are substantially reduced.
This will allow to use an smaller number of mock catalogues.
This kind of features and the performance shown by our method in this work, makes it a promising test to be applied on real data.


\section*{Acknowledgements}

This work has been partially supported by Consejo de Investigaciones Cient\'ificas y T\'ecnicas de la Rep\'ublica Argentina (CONICET) and the Secretar\'ia de Ciencia y T\'ecnica de la Universidad Nacional de C\'ordoba (SeCyT).
CMC acknowledges the hospitality of the Pontificia Universidad Cat\'olica de Chile (UC) and the Max Planck Institute for Extraterrestrial Physics (MPE), where part of this work has been done.
DJP acknowledges the financial support of the Agencia Nacional de Investigaci\'on Cient\'{\i}fica y T\'enica of Argentina (PICT-2862).
The academic exchanges and mobility of the authors during this work were partially supported by the Latin American Chinese European GALaxy Formation network (LACEGAL), and the DFG-CONICET bilateral cooperation program (Res. 1340/15).
REA acknowledges support of the European Research Council through grant number ERC-StG/716151, and of the Spanish Ministerio de Econom\'ia and Competitividad (MINECO) through grant number AYA2015-66211-C2-2.

Numerical calculations were performed at the computer clusters from the Centro de C\'omputo de Alto Desempe\~no de la Universidad Nacional de C\'ordoba (CCAD, http://ccad.unc.edu.ar).
Plots were made using the R software (https://www.r-project.org) and post-processed with Inkscape (https://inkscape.org).


\bibliographystyle{mnras}
\bibliography{references}


\appendix



\bsp	
\label{lastpage}
\end{document}